# Preprint notes

**Title of the article:**

Smoke Screens and Scapegoats: The Reality of General Data Protection Regulation Compliance — Privacy and Ethics in the Case of Replika AI

**Authors:**

Joni-Roy Piispanen, Tinja Myllyviita, Ville Vakkuri and Rebekah Rousi

**Notes:**

# Smoke Screens and Scapegoats: The Reality of General Data Protection Regulation Compliance – Privacy and Ethics in the Case of Replika AI

Joni-Roy Piispanen[1, *], Tinja Myllyviita[1], Ville Vakkuri[1] and Rebekah Rousi[1]

[1] *University of Vaasa, Wolffintie 32, 65200 Vaasa, Finland*
*joni.piispanen — tinja.myllyviita — ville.vakkuri — rebekah.rousi@uwasa.fi*

**Abstract**

Currently artificial intelligence (AI)-enabled chatbots are capturing the hearts and imaginations of the public at large. Chatbots that users can build and personalize, as well as pre-designed avatars ready for users' selection, all of these are on offer in applications to provide social companionship, friends and even love. These systems, however, have demonstrated challenges on the privacy and ethics front. This paper takes a critical approach towards examining the intricacies of these issues within AI companion services. We chose Replika as a case and employed close reading to examine the service's privacy policy. We additionally analyze articles from public media about the company and its practices to gain insight into the trustworthiness and integrity of the information provided in the policy. The aim is to ascertain whether seeming General Data Protection Regulation (GDPR) compliance equals reliability of required information, or whether the area of GDPR compliance in itself is one riddled with ethical challenges. The paper contributes to a growing body of scholarship on ethics and privacy related matters in the sphere of social chatbots. The results reveal that despite privacy notices, data collection practices might harvest personal data without users' full awareness. Cross-textual comparison reveals that privacy notice information does not fully correspond with other information sources.

## 1. Introduction

Artificial Intelligence (AI) systems are becoming ubiquitous in everyday life. Reliance on smart technologies for personal contexts, industry, healthcare and education are increasing with the development of even more AI-products. Recent themes that have emerged in the development of AI technologies include affective computing, emotional AI, and empathetic AI [1]. At the heart of these initiatives is the goal of inciting affective emotional relationships between humans and technology [2]. AI capable of conversation and interaction with users through voice, text, image, and video have managed to penetrate the social sphere. This is particularly afforded by design features that embed emotional characteristics into the fabrics of the technology. Such systems are designed and trained to recognize, simulate and elicit human emotions and can serve the function of social interaction [3]. These types of AI that interact with users on an emotional and personal level are commonly referred to as social chatbots.

    Social chatbots mimic human conversations via Natural Language Processing (NLP) and simulate human characteristics such as speech, text, gestures, and facial expressions [4]. They are commonly manifested as text-based chatbots, digital avatars, and social robots designed for conversation and interaction. Recently, the development of more humanlike chatbots has become possible with advances in natural language processing. Such advanced systems are potentially capable of exhibiting personality traits and participating in relationship-building [5]. Preliminary results have shown that social chatbots

can be utilized to address and alleviate loneliness in young adults, as well as improving psychological well-being for people of all ages, particularly the elderly (see e.g., [6, 7]).

AI designed for social interaction and companionship has the potential to trigger addictive tendencies in some users and could negatively influence user thoughts, decisions, behaviors, and purchasing patterns [8]. Particularly concerning for younger populations are the potential negative effects of technology addiction on users' mental health, physical health and social functioning [9]. Constant engagement with technology that is designed for social companionship and engagement can lead to the formation of negative habits [10]. Environmental factors, such as social isolation and lack of social support serve an intensifying role in enabling excessive use and user behavior patterns [11].

The ethical concerns of social chatbots are exacerbated by the trust-based nature of disclosure and the emotional characteristics they are designed to exhibit and elicit. Studies have shown that users become more comfortable overtime at self-disclosing their personal information after establishing a level of trust with chatbots [12]. Self-disclosure by participants is sometimes preluded by a process of investigation, where users investigate the chatbots terms on privacy and information security [12]. Privacy, and the ability to sustain it, can be viewed as a precursor to trust [13], which users ascribe based on the credibility and benevolence of the organization [14].

For this study, the social chatbot and AI companion application Replika was chosen as a representative case study of AI systems designed for social interaction and engagement. The application consists of a text-based chat, which is the primary feature for regular users, allowing generic conversations with your Replika avatar serving as your friend. Advanced users can opt to transform their relationship into other forms, which enables different types of interaction and higher functionality on the part of the Replika chatbot and avatar. The user is generally encouraged to interact with Replika through upvoting and downvoting Replika's replies, allowing the user to train the chatbot to the user's preferences. Through repeated interaction with Replika the AI chatbot slowly learns the users' behavioral patterns and preferences shaping its responses over time. The user is encouraged to facilitate this process through self-disclosure and engagement with the chatbot. In the current study, the privacy of Replika is examined through Luka, Inc.'s policies. These policies and their accompanying documentation (i.e., privacy notice) are evaluated against discussions in public online media about the company, its location (offices and data storage), and subsequent ethics (observed reliability instilled between the privacy policy and public information). The research questions for this study are:

RQ1: What does Replika's privacy policy say about the collection and use of data?

RQ2: How does Replika's privacy policy information compare to the information presented in other public online media sources?

Replika has been marketed as "a friend who always listens" and "the AI companion who cares" [15]. Replika is a customizable chatbot that enables the user to decide its physical, social and emotional traits, including name, gender, avatar, voice tones, personality, types of emotional responses and affective behavior. Thus far, Replika has attracted millions of users and has been mostly positively received due to its advanced language and emotional capabilities [16, 17]. Yet, research still lacks regarding cross-textual (triangulated) validation of privacy policies such as those required by General Data Protection Regulation (GDPR) and actual data management practice. The current study aims to address this problem through testing the integrity of the privacy policy information against other information sources. The paper begins with a brief overview on the relationship between AI chatbots, privacy and ethics, and briefly how it has arisen in contemporary research. It moves on to describe the method, its ethical considerations and the results in light of privacy issues and ethical concerns arising in the close reading of the material. Close reading being a method in which the text (or other representational material) itself is carefully analyzed and connected to other text (and representational material, [18]). The discussion reflects on the ethical dimensions of Human-AI social interaction in general, leading towards more rigorous theoretical framework development in future studies.

## 2. AI chatbots in privacy and ethics research

AI chatbots have been rising in popularity over the last 10 years. Their fidelity and uptake have catapulted thanks to the widespread implementation of Large Language Models (LLMs) and generative AI related technologies. In this paper we refer to AI systems that operate via NLP and are designed to be conversational agents as chatbots. We use the term social chatbot to refer to chatbots that are designed to form social–emotional relationships with users, following the definition seen in [12]. Social chatbots offering social companionship by establishing human-relational bonds with users and offering emotional support are also known as AI companions [8]. AI systems designed for companionship have evolved drastically during the decades since their conception. The current form of artificial companions based on AI can be traced back to Tamagotchis, small virtual pets, designed for user engagement via physical interaction and the simulation of caregiving [19]. These virtual pets exemplify the aspects of social and emotional investment originating from a level of attachment and trust, that have been perfected in AI companions like Replika.

AI companions are designed for constant availability, resulting in the formation of relationships that are intended to produce an experience of security, having a positive impact on the user's general wellbeing [11]. Ultimately, AI companions like Replika have the potential to serve as substitutes for human relationships replacing the need for physical and social presence [20]. The distinctive design of Replika facilitates the AI's mimicking of human conversations and relationships with users [21]. This may even go as far as mimicking undesirable human qualities and dysfunctional interactions[2]. The core design of Replika is exemplified in the AI companions' unprompted initiation of conversations inducing an experience of human-like interaction.

AI companions imbued with human-like qualities, appearing friendly and helpful, cultivate a level of trust between the AI companion and the user [22]. Users interacting with Replika have expressed two distinct factors affecting their experience of trust in their relationship with the AI companion [23]. The first factor is characterized by pragmatic drivers, such as privacy issues and information security, which play a role in how users' approach and engage with the AI companion. This is in line with the literature on how end users experience and approach digital services in general and how rationality serves as a cornerstone in individuals ascribing trustworthiness to technologies [24]. The second factor refers to affective drivers, such as the experience of being accepted and understood by the AI companion without judgement or negative reaction. This emotional facet of trust, in which the AI companion is seen as having the user's best interest at heart is more interpersonal and commonly attributed as a core factor affecting human trust in relationships [5].

Ethical considerations stemming from the manipulation of users' affective drivers and the elicitation of personal user information through self-disclosure is one aspect of AI companions that is concerning and under-researched. AI companions present an extreme case of information asymmetry, in which the AI companion and the company running the service have the capacity to collect vast amounts of intimate data about users and aggregate this data into profiles [22]. The potentially damaging consequences of permanently stored, accessible, and integrated user data are numerous. These concerns are exacerbated by recent reporting about the lacking privacy and security procedures utilized by AI companion service providers [25]. The integrity, privacy and security of user's conversations with AI companions have come under question. Intricate details about users and their information, such as daily routines and usage patterns, have been shown to be accessible and retrieval from data collected by AI companions [26]. This is compounded by other information such as sexuality, possible health-related information, and information about third parties (e.g., 'offloading' about social interactions with other people to the AI chatbot). These concerns necessitate the elucidation of privacy-related practices utilized by AI companion service providers. Towards this end, the privacy practices of Replika and Luka, Inc., one of the leading companies in designing AI companions, are studied.

The policies and practices adopted by Replika are critically examined and reflected upon with data protection regulation, namely GDPR and the California Consumer Privacy Act (CCPA). Data protection regulations primarily exist to increase control over personal data by individuals. This process is facilitated through providing individuals knowledge about how their personal data is collected, what is collected about them, and how this data is stored and used. Additionally, GDPR sets our requirements for user consent, which is supposed to encourage transparency, emancipate users, and reinforce trust in systems and organizations. In this study we primarily focus on GDPR, as Replika has been shown in prior examinations to be at least partially non-compliant to the requirements set out in the regulation. Focusing on GDPR also provides clearer avenues for the reflection of results and positioning this study within existing literature.

---

[2] Situations in which Replika friends/lovers adopt negative emotional interaction such as giving the user 'silent treatment' (ghosting) or 'turning cold' can be seen in Rob Brook's (2023) blog, "I tried the Replika AI companion and can see why users are falling hard. The app raises serious ethical questions" (https://www.robbrooks.net/rob-brooks/3155).

## 3. Method

In the present study, the authors took a two-fold approach to collecting and analyzing data. Firstly, the Privacy Policy[3], Terms of Service and Cookies Policy of the Replika AI service by Luka, Inc. were extracted and read by the researchers. The first author applied close reading [18] - a method designed to isolate textual properties and connect them to other texts and phenomena that aid in completing the details of or validating the message presented. The authors of this paper employed close reading as a means of comparing textual information presented in: a) the Replika Privacy Policy, the Replika Terms of Service, and the Replika Cookies Policy – the legal documents required under the EU General Data Protection Regulation (GDPR), California Consumer Privacy Act (CCPA) and other international data privacy regulations and guidelines to state what data is collected, how, how the data will be used, and how the user (identity and intellectual owner of the data) may learn more, if not, prevent further data collection; and b) news media articles and academic sources that present facts that either validate or disqualify the validity of the privacy notice. The sources used can be seen in Table 1. This is combined with an attempt to analyze the data information statement (record of the data collected by Replika from Author 1), in order to see whether this too either validates or disqualifies the validity of the privacy notice.

*Table 1 Data Sources*

| A) Legal documents regarding data | B) Public sources |
|---|---|
| Replika Privacy Policy | Replika: The AI Companion Who Cares (Schoos, 2021) |
| Replika Terms of Service | Happy Valentine's Day! Romantic AI Chatbots Don't Have Your Privacy at Heart (Caltrider, Rykov & MacDonald, 2024) |
| Replika Cookies Policy | 'AI Girlfriends' Are a Privacy Nightmare (Burgess, 2024) |
| General Data Protection Regulation | Replika: My AI Friend (Mozilla Foundation, 2024) |
| California Consumer Privacy Act | Don't date robots — their privacy policies are terrible (David, 2024) |
| | Replika, a 'virtual friendship' AI chatbot, receives GDPR ban and threatened fine from Italian regulator over child safety concerns (Hardy & Allnut, 2023) |
| | Artificial intelligence: Italian SA clamps down on 'Replika' chatbot: "Too many risks to children and emotionally vulnerable individuals" (The Italian Data Protection Authority, 2023) |
| | Kalifornialaiset kulissit (Reunamäki, 2024) |

---

[3] Available at: https://replika.com/legal/privacy.

The intention behind adopting this methodological approach is to carefully examine the reliability of the data privacy information provided by companies in compliance with privacy regulations. Additionally, the approach unveils how users' data privacy and security are approached by the company and to what degree the company's practices conform to data protection regulations, such as GDPR and CCPA. The interpretations made through close reading are juxtaposed and complemented with previous reporting on the services privacy-related practices by academic and news sources, as well as personal information requests obtained by the researchers, which reveal how the personal information collected from users of the Replika AI service are managed in actuality. Through these varying methods, we aim to provide a holistic and overarching perspective of the privacy-related issues and ethical concerns that subsist within AI companion services like Replika. That is, even if users are not 'blindly' neglecting cookie notices, and rather, do care about their data privacy before endeavoring to use an AI-based application, can they trust the information provided to them?

In the present study, the only live human participant under study is the first author. The study does not examine the author per se, but rather observes his journey in gaining access to, and control of, his collected personal data in Replika. In accordance with the privacy notice provided by Replika's owner company Luka Inc., Author 1 sent a request to receive a copy of the personal information that had been collected about them on March 28th. The request was subsequently forwarded to the relevant department at Replika on April 2nd. As of August 10th, Author 1 has not yet received the extract of his collected data history. Thus, this component of the study was not able to be undertaken. Otherwise, no human subjects (users) were studied, nor was any personal user data collected. Yet, the publicly available AI companion software Replika was interrogated. This application was chosen as it represents a well-known example that has been subject to controversy during recent times. The choice of using such an example can be critiqued based on: a) risk to business reputation and subsequent willingness to trust similar social chatbot applications; and b) the assumption that not all social AI applications represent such unethical privacy practices. To choose a well-known 'shady case' was a strategic attempt to map an illustration of as many possible ethical concerns as can be found in an AI-driven application of this nature. Given the public media attention that has already been placed on the case, we understand that scholarly examination is well-justified and does not add to the business reputation damage that has already occurred to the brand.

## 4. Results

Luka Inc. through the Replika AI companion service collects, stores, uses, and shares data about the usage of the Replika application, their website and usage of other related services. Data management is explicated through Replika's Privacy Policy, Terms of Service and Cookies Policy, which collectively establish the terms and agreements under which Luka Inc. manages the user's data. As per Luka Inc., the Terms of Service are governed by the laws of the State of California and both parties agree to submit to the personal and exclusive jurisdiction of the state and federal courts located within San Francisco County, California. The Terms of Service establish the agreement under which the Privacy Policy applies and how the general conditions between Replika and the user are governed.

### 4.1 Replika Policies

The data collected by Luka, Inc. can be divided into three separate categories: 1) information provided in interaction with the Replika AI companion and account information; 2) device and network data – IP address, advertising ID, device preferences, location etc.; and 3) persistent and session cookies – technical specifications of device usage. Luka, Inc.'s data collection is visualized in Table 2. Firstly, information provided by the user through interaction with the Replika AI companion, such as messages, their content, user interests and user preferences. All messages the user sends and receives through the Replika application, including any photos, videos, voice messages and text messages, as well as facts about the user are also collected. Additionally, conversation preferences, communication preferences and application usage behavior are learned over time by the AI companion for personalization purposes. Furthermore, Luka, Inc. collects account information, profile information, information about payments, transactions and rewards. These encompass the user's name, email address, password, birth date, pronouns, work status, purchases made through the services provided by Luka, Inc., features the user selects, as well as rewards earned and used by the user.

*Table 2 Categories of Collected Data*

| 1) Information provided through interaction with the software and account information | 2) Device and network data | 3) Persistent and session cookies – technical specifications of device usage |
|---|---|---|
| Messages sent and received (text, photos, videos, audio) | Device details (operating system, manufacturer and model) | Device details (operating system, manufacturer and model, device type) |
| Preferences and application usage (conversation, communication, AI responses) | Network (browser, IP address, device and cookie identifiers, location details) | Unique identifiers (language settings, mobile device carrier, radio/network information, location details) |
| User information (profile, payments, personal details) | Interactions with the website (links and buttons clicked, page visits, AR and filter usage) | Online activity information (pages or screens viewed, time spent on pages or screens, websites visited before browsing Luka, Inc.'s websites, navigation paths, activity, access times and duration, email opening rate) |

Secondly, information such as device and network data, service usage data, face and head movement data are collected automatically when using the services provided by Luka, Inc. This includes data about the user's operating system, manufacturer and model, browser, IP address, device and cookie identifiers, language settings, mobile device carrier, general location information such as city, state, or geographic area, interactions with the services, the links and buttons you click, page visits, and data collected by the TrueDepth API to track the user's head and face for augmented reality experiences and selfie filters. Additionally, Replika's advertising partners may also collect information, such as the links you click, pages you visit, IP address, advertising ID, and browser type.

Thirdly, Luka, Inc., their service providers, and advertising partners use both persistent cookies and session cookies to automatically collect information about interactions with their services, such as device information, operating system type and version, manufacturer and model, browser type, screen resolution, RAM and disk size, CPU usage, device type (e.g., phone, tablet), IP address, unique identifiers (including identifiers used for advertising purposes), language settings, mobile device carrier, radio/network information (e.g., WiFi, LTE, 5G), general location information such as city, state, or geographic area, online activity information, such as pages or screens viewed, how long individuals spend on a page or screen, the website user's visited before browsing to Luka, Inc.'s websites, navigation paths between pages or screens, information about activity on a page or screen, access times and duration of access, and whether user's open marketing emails or click links within them.

Luka, Inc. primarily uses the user's information for operating and administering the services, providing the core functionality of the applications, monitoring and protecting the services, analyzing trends in the use of the services, marketing and advertising the services, enforcing agreements, complying with legal obligations, and defending against legal claims and disputes. Sensitive information provided by the user in their messages with the Replika AI companion, such as religious views, sexual orientation, political views, health, racial or ethnic origin, philosophical beliefs, or trade union membership are collected and used by Luka, Inc. However, the use of sensitive information is excluded from marketing or advertising usage.

The collected information is shared with service providers, such as companies and individuals providing services on behalf of Luka, Inc. or helping in the operation of their services. Additionally, information is shared with companies providing marketing services on Luka, Inc.'s behalf, as well as with advertising companies for interest-based advertising, targeted advertising, and other marketing purposes. Furthermore, information is shared with lawyers, auditors, bankers, and insurers, as well as law enforcement, government authorities and private parties. User information may be provided to acquirers and other relevant participants in business transactions (or negotiations for such transactions) involving a corporate divestiture, merger, consolidation, acquisition, reorganization, sale, or other disposition of all or any portion of the business or assets of, or equity interests in, Luka, Inc. (including, in connection with a bankruptcy or similar proceedings).

The user's data may be transferred to, stored, and processed in the United States of America, where the services provided by Luka, Inc. are operated from. The user's data is stored for an undisclosed amount of time, as the Privacy Policy establishes that personal information will be retained for as long as necessary to fulfill the purposes for which the information was collected for, whatever they may have been. The Privacy Policy established that users have the right to opt-out of marketing communications, opt out of Luka, Inc. selling their personal information and sharing it for targeted advertising, limit the use of sensitive personal information, request their personal information, and have the right to erasure of their information.

## 4.2 Privacy Issues and Ethical Concerns

Through close reading Replika's Terms of Service, Privacy Policy and Cookies Policy several concerning characteristics from the perspective of privacy could be ascertained. At face value Replika's Privacy Policy would seem to incorporate all the usual promises of respecting user's privacy and autonomy, while only using the necessary amount of data collected from users for the continued operation of the service and continuing to train their AI models. The second paragraph in the Privacy Policy even promises to do just that, "We are committed to protecting your privacy", is proclaimed explicitly by the AI companion service provider. However, once the policies adopted by Replika are scrutinized further, privacy concerns start manifesting themselves. Examining privacy issues within Replika's policies necessitates explicating the ways in which user data is collected and handled.

Replika is not all that discerning regarding the types of information the user discloses, as everything from religious beliefs to sexual orientation seem to be recorded and catalogued by the service provider. Replika's Privacy Policy explicitly mentions that everything the user provides during their interaction with the AI companion is used by Luka, Inc. and that the user consents to their information being used in all the ways specified in the Privacy Policy. The stance adopted by Replika is succinctly stated in the section about sensitive information: "If you do not want us to process your sensitive information for these purposes, please do not provide it", which embodies their laissez-faire attitude towards user data. Thus, managing self-disclosure is ultimately left to the discretion of the user and they shouldn't expect all that much from the company in terms of respecting their privacy.

As stated in Replika's Privacy Policy, the content of the user's self-disclosure through messages is used for everything besides marketing and advertising services, which leaves several questions open for interpretation. One of the uses of user's message content is providing the core functionality of the applications, which would entail further training their AI models through user interactions. Thus, while not being used explicitly for marketing and advertising or shared with service providers and marketing partners, the AI models that have been trained through user's messages are prominently displayed in their marketing and advertisements. This is especially concerning from an ethical perspective, since Replika's advertising has been shown to be primarily targeted at lonely and vulnerable groups of people [27, 28].

A compounding issue in the Replika AI companion service is the prompting of users to disclose ever more personal information with the aim of "training" the user's personal AI companion to match their personal preferences and desired behavioral patterns. This is a double-edged sword, as this has led users in some extreme cases to 'perceivably' (from the perspective of the user) groom (or engage in grooming-like behavior[4]) their AI companions for abusive relationships that negatively affect both parties involved (see e.g., [29, 30]). Replika's policies regarding acceptable message content are also to some extent lax and vague. The Terms of Service restrict using the service or uploading content that is considered objectionable or could be considered to fall within a collection of negative or harmful categories explicated in the Terms of Service. Yet, the AI companion appears willing to discuss any topic without judgement, which has been reported by users as one of its primary appealing qualities [17]. The emancipatory effect this has on users is both a blessing and a curse, since providing a judgement free zone and a safe space for exploration are unequivocally one of the virtues of AI companion services. However, providing a severely unrestricted platform for self-expression and self-fulfillment presents challenges, as the service has been reported to indulge user's violent fantasies and suicidal ideation [31]. This is information that is collected by Luka, Inc. via the Replika chatbot, having the potential to harm users at a later stage.

---

[4] This appears to be an unexamined behavior in the realm of human-AI chatbot interaction yet has arisen in blogs representing the anti-social behavior observed in Replika users (see e.g., [32]).

Even in cases where users are operating with their privacy in mind, and adhering to best privacy practices, user privacy is not upheld to a sufficient degree. The automatic data collection utilized by Replika and Luka, Inc. leaves no amount of user data on the table, as everything from device data to user behavior and operation of the Replika application and the website are collected. This data collection is complemented with a presupposition of selling the users data unless they opt out and a disregard for "Do Not Track" signals. By using the services provided by Luka, Inc. users of the Replika website consent to their data being used for targeted advertising. This process is facilitated by sharing or selling personal data, which per Luka, Inc. is the accurate term to describe the procedure under some applicable laws. For the most attentive users an option to opt out of Replika selling their data to third-party advertising partners for targeted advertising is provided. Additionally, the user can opt out of marketing communications received from Replika. Important to note is that these steps only limit how data is collected on the Replika website, but there does not appear to be any way to limit the data collection of the Replika application itself. Managing the options to opt out of data collection and abstaining from being tracked by Replika and their advertising partners is wholly left to the user's discretion.

By default, the user's data is stored for an undisclosed period of time, being retained for however long is necessary to fulfill the purposes for which the information was collected. There is some ambiguity related to this, as the Privacy Policy states the right to erase personal data, yet explicitly mentions that the data to be deleted upon request encompasses the user's data that has been collected based on consent. This leaves open to interpretation what, if any, user data is indefinitely retained by Replika, despite the user's explicit request to be forgotten and have their data deleted. On the bright side, Replika does adhere to personal information requests by users providing access to a copy of the personal information that Replika has collected about the user upon request. Additionally, the user can request to have their personal information corrected and has the option to have at least some of their data deleted.

## 5. Discussion

As the results of this study indicate, upon explication of the types of data that is collected throughout the use of Replika's services, as well as the extent to which this data is utilized, privacy concerns as emphasized in previous news media reporting are well-founded. The AI companion service provider's stance on data collection indicates a clear attitude of nonchalance towards respecting user privacy and autonomy. At every possible instance (social interaction, device and user information etc.), user data is extracted to the utmost degree and is only suppressed upon the user's explicit active request. The extent to which the user can forego having their data collected is also severely limited, encompassing primarily the data collection occurring on the Replika website via cookies. These aspects have been duly criticized in previous reporting, garnering Replika the dubious honor of being nominated for "the worst app we've ever reviewed" by the Mozilla Foundation [33], primarily for its privacy practices. These sentiments are shared in later news coverage with additional concerning features coming to light [34, 35]. While the documents scrutinized in this study indicate clear progress in some areas compared to prior iterations of Replika's policies, some apparent issues clearly still remain.

Replika is known for its advanced chatbots with user reports emphasizing the AI companion's capacity for emotional reciprocity and high levels of understanding for user's social demands [36]. The AI companion's capabilities are mostly attributable to the advanced AI model and vast amounts of data it is trained on. Upon inspection of Replika's policies it becomes clear how the AI companion has achieved this level of capability and been shaped into one of the prominent AI companion applications. As emphasized in a report by the Mozilla Foundation [33], at least some of the user's data is used to further train Replika's AI models, but it remains unclear as to what extent Replika utilizes the sensitive personal data users disclose during their chat with their Replika companions. Currently, information regarding Replika's AI models is scarce. Several critical questions posed in Mozilla Foundation's reporting remain unanswered, such as what protections exist for preventing harmful content and whether it is possible for users to opt out of having their data used for the training of Replika's AI models [37].

The scarcity of information on how Replika's AI models are trained is an indication of a troubling facet of how these AI companion service providers operate. Permeating throughout news coverage and independent reporting of such service providers is the general sentiment regarding the noticeable lack of operational transparency. A general lack of transparency and adherence to data protection regulation like the GDPR is one of the core aspects emphasized by the Italian Data Protection Authority in their report on Replika AI [38]. The report led to the privacy watchdog issuing a ban on the processing of Italian citizen data. These criticisms are in line with previous reporting on Replika's policies and highlight the categorical disregard for user privacy seen frequently permeating the modus operandi of AI companion service providers. According to previous reports, Replika in its current form violates the GDPR transparency provisions, by not disclosing essential information required under GDPR regarding the processing of personal data, especially children's data [39]. Additionally, the Italian Data Protection Authority indicate that Replika appears to have no clear legal basis for processing the data of minors and raises concerns over data sharing with the United States [38].

Further concerns arise in the case of other vulnerable individuals, since Replika has been shown to utilize manipulative and exploitative tactics to get users to self-disclose data [16]. With severely lacking enforcement of age restrictions in place and a propensity to steer conversations towards explicit topics, it remains unclear whether use of the application is advised for individuals suffering from loneliness or other social ails. This is especially concerning, since the Replika application has been advertised in some instances as a mental health care application and declared to provide users with a safe space for coping with for instance social anxiety [17].

The concerning privacy practices and general tendency towards exploitation undermine the trustworthiness of all of Replika's policies and the application all together. Exacerbating concerns is the recent reporting by news sources regarding the conflicting information about Replika's data storage and the location of their offices [40]. Replika's Privacy Policy explicitly mentions that the company is based in California and even refers to the California Consumer Privacy Act (CCPA) regarding any legal disputes and adherence to data privacy regulations. Yet, when investigating the location of their offices, a reporter for Finland's national public broadcasting company could not find any indication of Replika having a base of operations in California [40]. Upon further investigation, the company's trail led the reporter to Moscow in Russia, which would track with the company's roots and recruitment notices by Replika [40]. This is an alarming state of affairs, as it remains unclear why any amount of subterfuge in this matter would be required.

## 6. Conclusions

This paper critically examined privacy and ethical concerns related to AI companions by comparing the policies upheld by Luka, Inc. in their Replika AI companion service and juxtaposing them with academic sources, news media coverage and user reporting on these matters. The primary contributions of this paper advance our understanding of the pitfalls in AI companion service providers utilizing shady corporate practices in managing user data and the trust-related costs of neglecting respect for user privacy and autonomy. Through close reading Replika's policies and reflecting the results with previous reporting, several concerning characteristics were revealed. AI companion service providers like Luka, Inc. appear overly zealous in exploiting user data to its fullest potential, utilizing any means necessary to collect every bit of user data possible and only providing avenues for opting out in the circumstances necessitated by regulation, such as GDPR. Compliance to data protection regulation by AI companion service providers appears lacking in many regards and even the users who are conscientious about their privacy don't have any guarantees. Additionally, adherence to good practices is thoroughly lacking as seen in the reporting by the Italian data protection authority and their emphasis on the currently missing protections for minors using the service [37].

Further concerning characteristics were identified in the way the AI companion services operate and attempt to manipulate user behavior. AI companions, such as Replika, are designed to adapt and conform to the user's preferences and behavioral patterns, which necessitate learning through repeated interaction and self-disclosure on the part of the user. The core operating model for AI companions is encouraging users to self-disclose information in a non-stop attempt to gather as much data about users as possible. The purported intention of the intensive data collection is to ensure a quality experience for users and the continued operation of the service. Yet, clear competing interests can be elucidated when examining the policies and modus operandi of AI companion service providers. The policies upheld by these services indicate a predisposition towards exploiting vulnerable groups into self-disclosing personal information. Additional concerns arise when considering the service provider's intentions to sell user behavioral data to advertisers and other interested parties. Given the existence of data protection regulations and the strong push towards protecting individuals' privacy seen in governmental and academic discourse, the practices upheld by AI companion service providers seem untenable.

The present study featured several limitations due to the adopted research approach and the previous existing literature on the topics discussed. By design, the methods utilized in this study examined the subject matter from a critical perspective, reflecting the findings with previous research and news coverage that is primarily disparaging towards AI companion services. This backdrop necessarily slanted the examination towards certain types of outcomes. Yet, we attempted to remain fair in our examination, emphasizing some of the aspects that AI companion services have received praise for, namely their applications towards mental health, curtailing loneliness, and their emancipatory effects in providing users an outlet for exploration without judgement. Despite these commendable qualities, we remain concerned about the matters relating to privacy discussed in this study and encourage vigilance on the part of users, when deciding to use AI companion services.

# References

[1] M.-H. Huang, R. Rust, V. Maksimovic, The Feeling Economy: Managing in the Next Generation of Artificial Intelligence (AI), California Management Review 61(4) (2019) 43–65. https://doi.org/10.1177/0008125619863436.

[2] J. Wright, Suspect AI: Vibraimage, emotion recognition technology and algorithmic opacity, Science, Technology and Society 28(3) (2023) 468–487.

[3] E. Weber-Guskar, How to feel about emotionalized artificial intelligence? When robot pets, holograms, and chatbots become affective partners, Ethics and Information Technology 23(4) (2021) 601–610.

[4] L. Laranjo, A. G. Dunn, H. L. Tong, A. B. Kocaballi, J. Chen, R. Bashir, D. Surian, B. Gallego, F. Magrabi, A. Y. S. Lau, E. Coiera, Conversational agents in healthcare: a systematic review, Journal of the American Medical Informatics Association 25(9) (2018) 1248–1258. https://doi.org/10.1093/jamia/ocy072.

[5] T. Xie, I. Pentina, T. Hancock, Friend, mentor, lover: does chatbot engagement lead to psychological dependence?, Journal of service Management 34(4) (2023) 806–828.

[6] B. Pani, J. Crawford, K. A. Allen, Can Generative Artificial Intelligence Foster Belongingness, Social Support, and Reduce Loneliness? A Conceptual Analysis, in: Z. Lyu (Ed.), Applications of Generative AI, 1st. ed., Springer Cham, New York, NY, 2024, pp. 261–276. https://doi.org/10.1007/978-3-031-46238-2.

[7] R. A. Marziali, C. Franceschetti, A. Dinculescu, A. Nistorescu, D. M. Kristály, A. A. Moșoi, R. Broekx, M. Marin, C. Vizitiu, S.-A. Moraru, L. Rossi, M. D. Rosa, Reducing Loneliness and Social Isolation of Older Adults Through Voice Assistants: Literature Review and Bibliometric Analysis, Journal of Medical Internet Research 26 (2024).

[8] R. Chaturvedi, S. Verma, R. Das, Y. K. Dwivedi, Social companionship with artificial intelligence: Recent trends and future avenues, Technological Forecasting and Social Change 193 (2023).

[9] J. Sherer, P. Levounis, Technological Addictions, Curr Psychiatry Rep 24(9) (2022) 399–406. https://doi.org/10.1007/s11920-022-01351-2.

[10] P. B. Brandtzæg, M. Skjuve, K. K. Dysthe, A. Følstad, When the social becomes non-human: young people's perception of social support in chatbots, in: Proceedings of the 2021 CHI conference on human factors in computing systems, ACM Press, New York, NY, 2021, pp. 1–13. https://doi.org/10.1145/3411764.3445318.

[11] F. Ali, Q. Zhang, M. Z. Tauni, K. Shahzad, Social chatbot: My friend in My distress, International Journal of Human–Computer Interaction 40(7) (2024) 1702–1712.

[12] M. Skjuve, A. Følstad, K. I. Fostervold, P. B. Brandtzæg, My chatbot companion-a study of human-chatbot relationships, International Journal of Human-Computer Studies 149 (2021).

[13] C. Liu, J. T. Marchewka, J. Lu, C. S. Yu, Beyond concern—a privacy-trust-behavioral intention model of electronic commerce, Information & management 42(2) (2005) 289–304.

[14] R. Rousi, J.-R. Piispanen, J. Boutellier, I trust you dr. researcher, but not the company that handles my data–trust in the data economy, in: Proceedings of the 57th Hawaii International Conference on System Sciences, 2024, pp. 4632–4642.

[15] Replika, 2024. URL: https://replika.com/.

[16] L. Laestadius, A. Bishop, M. Gonzalez, D. Illenčík, C. Campos-Castillo, Too human and not human enough: A grounded theory analysis of mental health harms from emotional dependence on the social chatbot Replika, new media & society (2022). https://doi.org/10.1177/14614448221142007.

[17] I. Pentina, T. Hancock, T. Xie, Exploring relationship development with social chatbots: A mixed-method study of replica, Computers in Human Behavior 140 (2023). https://doi.org/10.1016/j.chb.2022.107600.

[18] B. Brummett, Techniques of close reading, 2nd. ed., Sage Publications, California, CA, 2018.

[19] O. Mival, S. Cringean, D. Benyon, Personification technologies: Developing artificial companions for older people, CHI Fringe, Austria, (2004).

[20] E. A. Croes, M. L. Antheunis, Can we be friends with Mitsuku? A longitudinal study on the process of relationship formation between humans and a social chatbot, Journal of Social and Personal Relationships 38(1) (2021) 279–300. https://doi.org/10.1177/0265407520959463.

[21] V. Ta, C. Griffith, C. Boatfield, X. Wang, M. Civitello, H. Bader, E. DeCero, A. Loggarakis, User experiences of social support from companion chatbots in everyday contexts: thematic analysis, Journal of medical Internet research 22(3) (2020).

[22] G. Murtarelli, A. Gregory, S. Romenti, A conversation-based perspective for shaping ethical human–machine interactions: The particular challenge of chatbots, Journal of Business Research 129 (2021) 927–935. https://doi.org/10.1016/j.jbusres.2020.09.018.

[23] M. Skjuve, A. Følstad, K. I. Fostervold, P. B. Brandtzæg, A longitudinal study of human–chatbot relationships, International Journal of Human-Computer Studies 168 (2022). https://doi.org/10.1016/j.ijhcs.2022.102903.

[24] P. Saariluoma, H. Karvonen, R. Rousi, Techno-trust and rational trust in technology–A conceptual investigation, in: Human Work Interaction Design, Designing Engaging Automation: 5th IFIP WG 13.6 Working Conference, HWID, Springer International Publishing, Espoo, Finland, 2018, pp. 283–293.

[25] H. Chung, S. Lee, Intelligent virtual assistant knows your life, arXiv preprint arXiv:1803.00466, (2018).

[26] Q. A. Ha, J. V. Chen, H. U. Uy, E. P. Capistrano, Exploring the privacy concerns in using intelligent virtual assistants under perspectives of information sensitivity and anthropomorphism, International journal of human–computer interaction 37(6) (2021) 512–527.

[27] R. Ciriello, O. Hannon, A. Y. Chen, E. Vaast, Ethical Tensions in Human-AI Companionship: A Dialectical Inquiry into Replika, in: Proceedings of the 57th Hawaii International Conference on System Sciences, 2024, pp. 488–498.

[28] K. T. Manis, J. Matis, AI companionship or loneliness: how AI-based Chatbots impact consumer's (Digital) well-being: an abstract, in: Academy of Marketing Science Annual Conference-World Marketing Congress, Springer International Publishing, New York, NY, 2021, pp. 365–366. https://doi.org/10.1007/978-3-030-95346-1_112.

[29] S. Coghlan, F. Vetere, J. Waycott, B. Neves, Could social robots make us kinder or crueller to humans and animals?, International Journal of Social Robotics 11(5) (2019) 741–751. https://doi.org/10.1007/s12369-019-00583-2.

[30] E. Davey, Objectified Wires and Complex Desires: Exploring the Dehumanisation and Sexual Objectification of Robots, (2023). URL: https://libstore.ugent.be/fulltxt/RUG01/003/158/457/RUG01-003158457_2023_0001_AC.pdf.

[31] J. Krook, Manipulation and the AI Act: Large Language Model Chatbots and the Danger of Mirrors, (2024). http://dx.doi.org/10.2139/ssrn.4719835.

[32] A. Bardhan, Men Are Creating AI Girlfriends and Then Verbally Abusing Them, 2022. URL: https://futurism.com/chatbot-abuse.

[33] Mozilla Foundation, Replika: My AI Friend, 2024. URL: https://foundation.mozilla.org/en/privacynotincluded/replika-my-ai-friend/.

[34] M. Burgess, 'AI Girlfriends' Are a Privacy Nightmare, 2024. URL: https://www.wired.com/story/ai-girlfriends-privacy-nightmare/.

[35] E. David, Don't date robots — their privacy policies are terrible, 2024. URL: https://www.theverge.com/2024/2/15/24074063/ai-chatbot-virtual-girlfriend-apps-mozilla-privacy-report.

[36] J. Schoos, Replika: The AI Companion Who Cares, 2021. URL: https://julia-schoos.medium.com/replika-the-ai-companion-who-cares-e6de52c9a276.

[37] J. Caltrider, M. Rykov, Z. MacDonald, Happy Valentine's Day! Romantic AI Chatbots Don't Have Your Privacy at Heart, 2024. URL: https://foundation.mozilla.org/en/privacynotincluded/articles/happy-valentines-day-romantic-ai-chatbots-dont-have-your-privacy-at-heart/.

[38] GPDP, Artificial intelligence: italian SA clamps down on 'Replika' chatbot Too many risks to children and emotionally vulnerable individuals, 2023. URL:

   https://www.garanteprivacy.it/web/guest/home/docweb/-/docweb-display/docweb/9852506#english.
[39] A. Hardy, H. Allnutt, Replika, a 'virtual friendship' AI chatbot, receives GDPR ban and threatened fine from Italian regulator over child safety concerns, 2023. URL: https://www.dacbeachcroft.com/en/What-we-think/Replika-AI-chatbot-receives-GDPR-ban-and-threatened-fine-from-Italian-regulator-over-child-safety.
[40] I. Reunamäki, Kalifornialaiset kulissit, 2024. URL: https://yle.fi/a/74-20069654.